\documentclass{webofc}

\usepackage[varg]{txfonts}   
\usepackage{hyperref}
\usepackage{url}
\hypersetup{colorlinks=true,citecolor=blue,urlcolor=blue,linkcolor=blue}
\newcommand{\PbPb}          {\mbox{Pb--Pb}\xspace}

\newcommand{\snn}           {\ensuremath{\sqrt{s_{\mathrm{NN}}}}\xspace}
\newcommand{\vn}            {\ensuremath{v_{\rm n}}\xspace}
\newcommand{\vtwo}          {\ensuremath{v_{\rm 2}}\xspace}

\newcommand{\pt}            {\ensuremath{p_{\rm T}}\xspace}
\newcommand{\RAA}           {\ensuremath{R_\mathrm{AA}}\xspace}
\newcommand{\TeV}           {\ensuremath{\mathrm{TeV}}\xspace}
\newcommand{\GeVc}          {\ensuremath{\mathrm{GeV}/c}\xspace}
\newcommand{\nb}            {\ensuremath{\mathrm{nb}^{-1}}\xspace}
\newcommand{\pip}           {\ensuremath{\mathrm{\pi^{+}}}\xspace}
\newcommand{\Dzero}         {\ensuremath{\mathrm{D^0}}\xspace}
\newcommand{\Dplus}         {\ensuremath{\mathrm{D^+}}\xspace}
\newcommand{\Ds}            {\ensuremath{\mathrm{D_s^+}}\xspace}

\newcommand{\Lc}            {\ensuremath{\Lambda_\mathrm{c}^+}\xspace}
\newcommand{\Jpsi}          {\ensuremath{\mathrm{J/\psi}}\xspace}

\begin{document}
\title{Investigating charm-quark dynamics in the QGP via the charm-hadron elliptic flow in \PbPb collisions with ALICE}
%
%
\author{
\firstname{Chuntai} \lastname{Wu}\inst{1,2}\fnsep\thanks{\email{chuntai.wu@cern.ch}} 
}
\institute{Padova University
\and
           Central China Normal University 
          }
\abstract{
In these proceedings, we report the measurement of the elliptic flow (\vtwo) of prompt and non-prompt \Dzero, prompt \Dplus and \Ds mesons, and, for the first time ever at the LHC, prompt and non-prompt \Lc baryons in \PbPb collisions at $\snn = 5.36\ \TeV$ collected by the ALICE experiment during the LHC Run 3. The analysis is performed at midrapidity ($|y| < 0.8$) in different centrality classes (30--40\%, 40--50\%, and 60--80\% for \Dzero, \Dplus, and \Ds; 30--50\% for \Lc). The candidates are reconstructed from hadronic decay channels, and the \vtwo is extracted via the scalar-product method. The measurements are compared to light-flavor results and several predictions of models that differently describe heavy-quark transport and hadronization processes in the QGP.
}
\maketitle
\section{Introduction}
\label{intro}
Heavy quarks (charm and beauty) are excellent probes for investigating the properties of the quark--gluon plasma (QGP) generated in ultra-relativistic heavy-ion collisions. Their participation in the collective motion of the medium can be studied by measuring the elliptic flow of charm hadrons. In non-central heavy-ion collisions, the elliptic flow originates mainly from the initial-state spatial asymmetry~\cite{Liu:2012ax}. The elliptic flow \vtwo is the second-order coefficient of the Fourier decomposition of the particle azimuthal distribution. The harmonic coefficients, which quantify the flow anisotropy (\vn), are calculated as $\vn =  \langle\cos[n(\phi-\Psi_{n})]\rangle$, where $\phi$ is the particle’s azimuthal angle and $\Psi_{n}$ denotes the symmetry-plane angle for the $n^{\mathrm{th}}$ harmonic~\cite{Poskanzer:1998yz}. These measurements provide essential inputs to constrain theoretical models describing the charm-quark transport in the QGP, being sensitive to charm thermalization, heavy-quark diffusion coefficients, path-length dependence of parton energy loss, initial-state spatial anisotropy, and hadronization mechanisms. In particular, comparisons between charm meson and baryon \vtwo can provide further insights into medium-induced phenomena, such as the interplay of radial flow and the charm-quark hadronization via coalescence. Comparisons between charm and beauty also allow exploration of the mass dependence of relaxation times.

\section{Data sample and analysis procedure}
\label{DA}
In these proceedings, we present measurements of the elliptic flow \vtwo of \Dzero, \Dplus, and \Ds mesons, and, for the first time ever at LHC, \Lc baryons, in \PbPb collisions at $\snn =\ 5.36\ \TeV$, within the rapidity interval $|y| < 0.8$. The analysis is based on the data samples collected with the ALICE detector in 2023. After the LHC Long Shutdown 2, the ALICE detector underwent a major upgrade, as detailed in Ref.~\cite{ALICE:2023udb}. The integrated luminosity of this minimum-bias data sample in 2023 reached approximately 1.5 \nb. The \vtwo of prompt \Dzero, \Dplus, and \Ds mesons is measured in the 30--50\%, 30--40\%, 40--50\%, and 60--80\% centrality classes. The \vtwo of the prompt and non-prompt \Lc baryons, as well as the non-prompt \Dzero meson, is measured in the 30--50\% centrality class. 
The candidates are reconstructed by combining couples or triplets of tracks that pass some quality cuts, with the correct charge combination. To reduce the combinatorial background and separate the prompt and non-prompt contributions, a multi-class Machine Learning (ML) model, implemented as Boosted Decision Trees (BDT)~\cite{Chen:2016btl}, was trained, following the training strategy described in Ref.~\cite{ALICE:2023gjj}. Signal candidates were selected using the BDT-based selections. The inclusive \vtwo was extracted using simultaneous fits to the invariant mass spectrum and the \vtwo as a function of mass, and the feed-down fraction was estimated using a data-driven method~\cite{ALICE:2023gjj}. A linear fit to the inclusive \vtwo as a function of the feed-down fraction was used to evaluate the prompt and non-prompt \vtwo components.

\section{Charm meson \vtwo}
\label{Rmeson}
\begin{figure}[h]
\centering
\setlength{\abovecaptionskip}{6pt}
\includegraphics[width=0.40\columnwidth,clip]{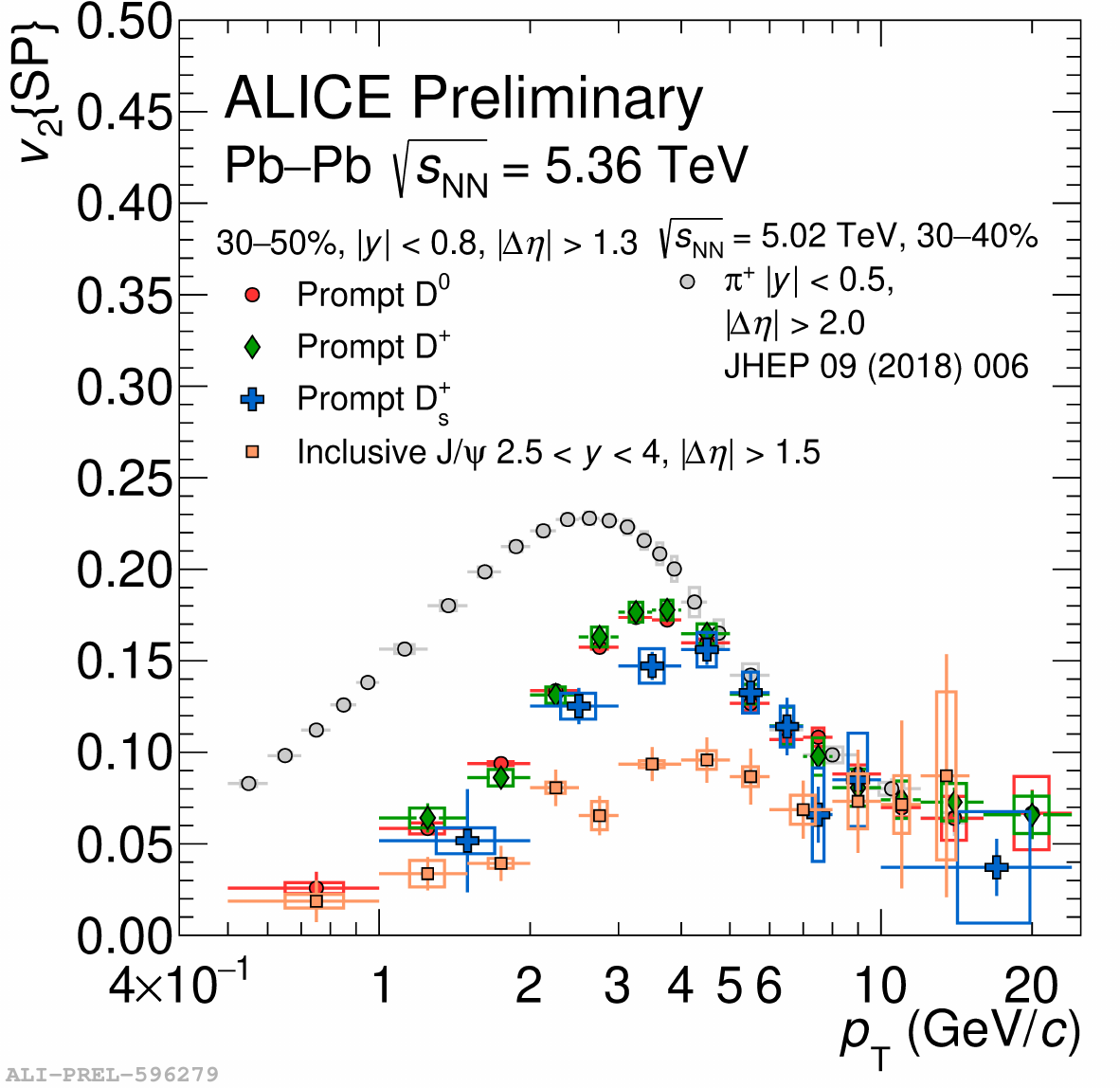}
\includegraphics[width=0.59\textwidth,clip]{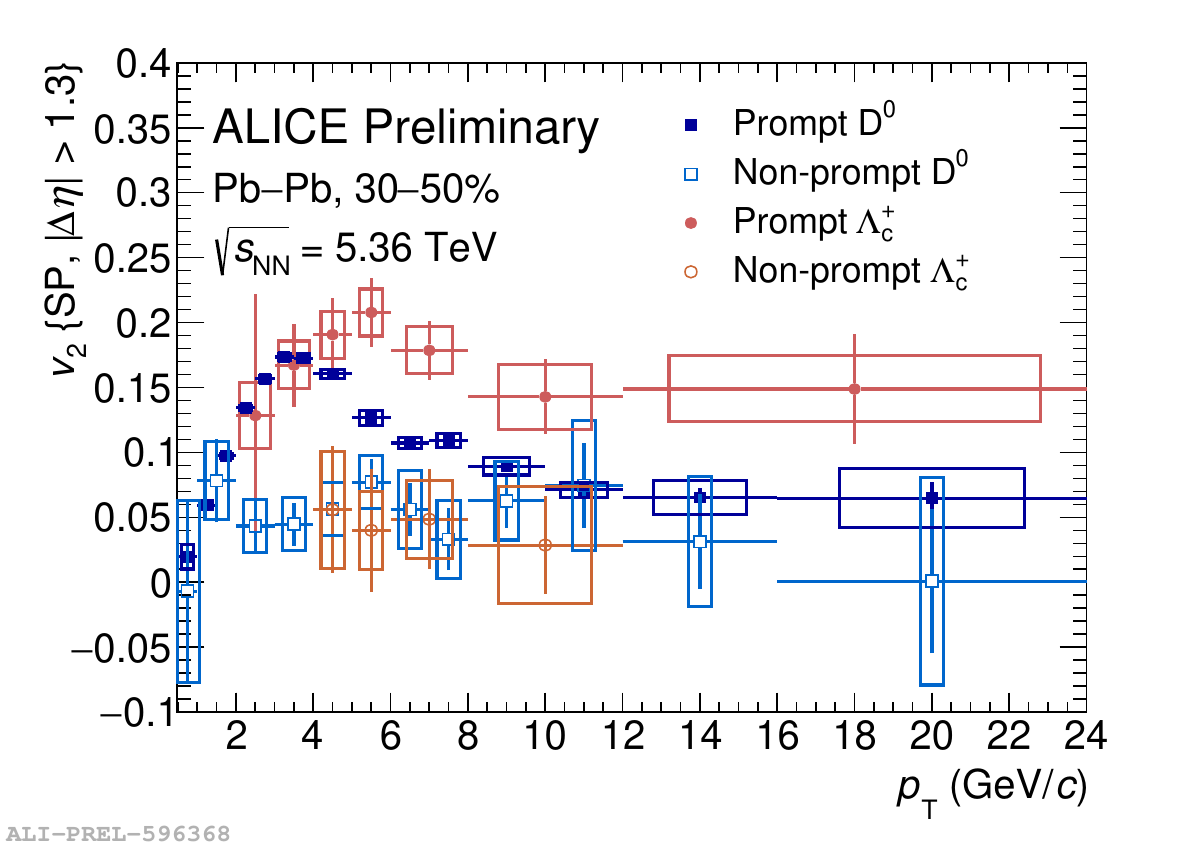}
\caption{Left: elliptic flow (\vtwo) of prompt \Dzero, \Dplus, and \Ds mesons in the 30–50\% centrality class is compared to that of inclusive \Jpsi and \pip~\cite{ALICE:2018yph}. Right: measured \vtwo of prompt and non-prompt \Lc baryons and \Dzero mesons in the 30-50\% centrality interval.  
}
\vspace{-15pt}
\label{Dmesons_v2}       
\end{figure}

\begin{figure*}[htbp]
\centering
\setlength{\abovecaptionskip}{6pt}
\includegraphics[width=0.9\textwidth,clip]{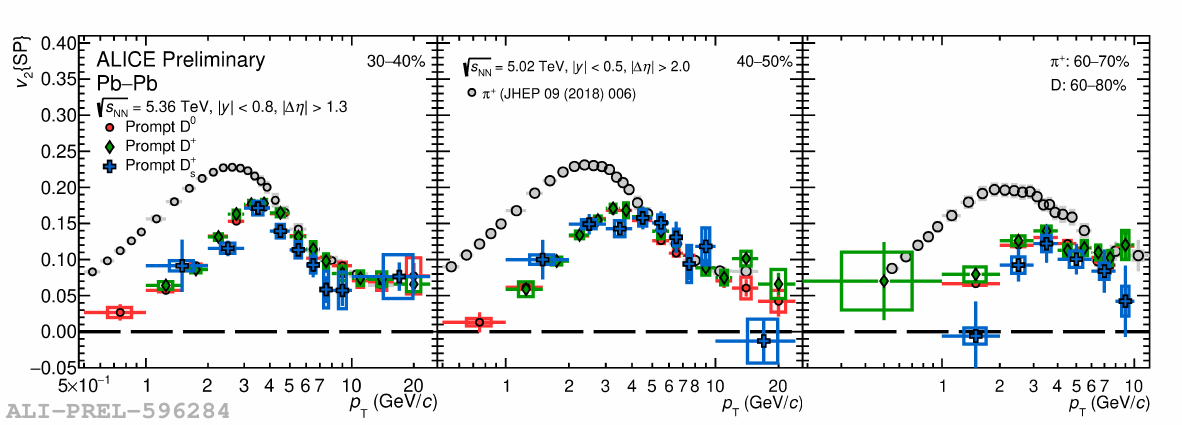}
\caption{Measured \vtwo of prompt \Dzero, \Dplus, and \Ds mesons, in 30-40\% (left), 40-50\% (middle), and 60-80\% (right), compared to that of \pip.}
\vspace{-15pt}
\label{D_v2_cent}       
\end{figure*}

Figure~\ref{Dmesons_v2} (left) shows the \vtwo of prompt \Dzero, \Dplus, and \Ds mesons in the 30-50\% centrality interval, compared to that of inclusive \Jpsi in the same centrality interval and of \pip~\cite{ALICE:2018yph} in 30-40\%. This is the first measurement of the prompt \Dzero meson \vtwo at the transverse momentum \pt < 1 \GeVc. Overall, the \vtwo of \Dplus mesons shows a good agreement with that of \Dzero mesons. At \pt < 4~\GeVc, a mass ordering is observed when comparing \vtwo of the heavy-flavor and light-flavor hadrons. The \Ds meson \vtwo is close to the non-strange D meson \vtwo, but there is a hint that it is smaller, with a deviation of 2.1 $\sigma$. 
In the interval 4 < \pt < 8~\GeVc, the prompt D-meson \vtwo is compatible with the \vtwo of \pip and both have a decreasing trend. The \Jpsi \vtwo is smaller than that of D mesons for \pt < 10~\GeVc. Above 10 \GeVc, the \vtwo measurements of the different particle species are compatible.  
The common \vtwo values of the different particle species for \pt > 10 \GeVc can indicate a similar effect from the path-dependent parton energy loss in the QGP. 
As shown in Fig.~\ref{D_v2_cent}, the \vtwo of prompt \Dzero, \Dplus, and \Ds mesons in 30-40\%, 40-50\%, and 60-80\% centrality intervals is compared with that of \pip~\cite{ALICE:2018yph}. A significant modulation of \vtwo as a function of \pt is observed in the 30-40\% and 40-50\% centrality classes, while a flatter \pt trend is observed in 60-80\%. This centrality-dependent behavior could provide crucial constraints to the transport properties of QGP, as well as to the interplay of collision eccentricity and QGP size and lifetime in determining the final \vtwo.

\section{Prompt and non-prompt \Lc-baryon \vtwo}
\label{Rbaryon}
\begin{figure}[h]
\centering
\setlength{\abovecaptionskip}{6pt}
\includegraphics[width=0.65\textwidth,clip]{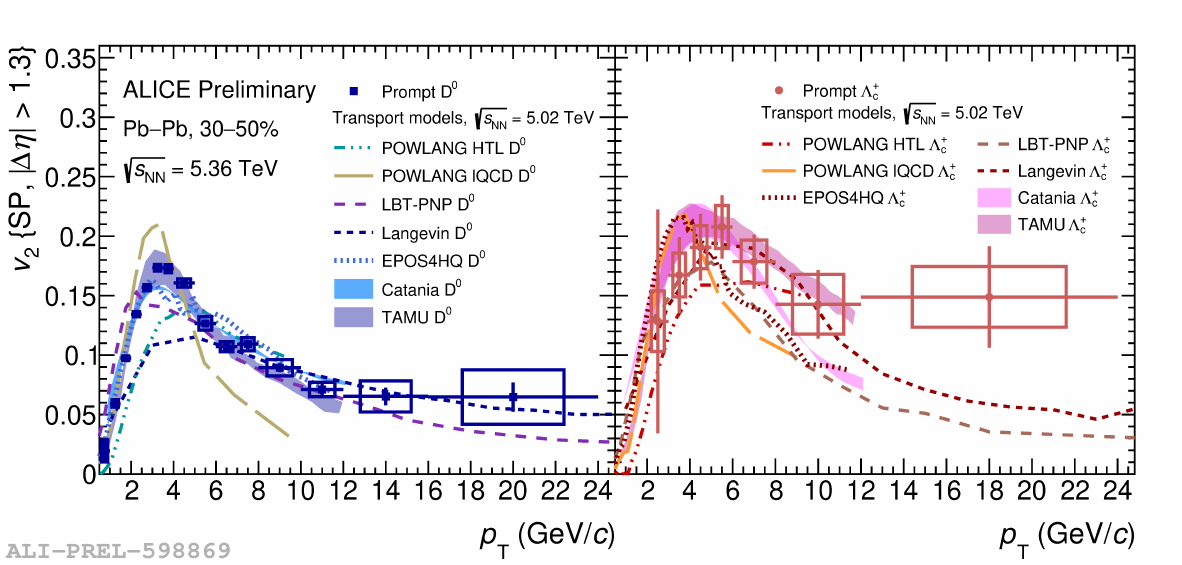}
\caption{Comparison of measured \vtwo of prompt \Lc baryons and \Dzero mesons with model predictions.}
\vspace{-15pt}

\label{Lc_v2}       
\end{figure}

The \Lc baryon, composed of one charm quark and two light quarks, is expected to be more sensitive to the coalescence mechanism in QGP. As shown in Fig.~\ref{Dmesons_v2} (right), for the first time ever at LHC, the \vtwo of prompt and non-prompt \Lc baryons are measured in heavy-ion collisions. At \pt < 4 \GeVc, the \vtwo of prompt \Lc baryons is compatible with that of prompt \Dzero mesons within uncertainties. At \pt > 4 \GeVc, the prompt \Lc-baryon \vtwo is larger than that of \Dzero mesons with a significance of 3.6 $\sigma$, which provides the first evidence of baryon-meson splitting in the charm sector. The non-prompt \Lc-baryon \vtwo is consistent with that of non-prompt \Dzero mesons within uncertainties, and both are smaller than the prompt \vtwo. This difference can be ascribed to the larger mass of beauty quarks, which are less influenced by the collective motion and have a longer relaxation time. Figure~\ref{Lc_v2} shows the comparison of the measured \vtwo of prompt \Dzero mesons and \Lc baryons with several model predictions~\cite{He:2019vgs, Li:2020umn, Cassing:2009vt, Sambataro:2024mkr, Beraudo:2023nlq, Xing:2021xwc}, which incorporate different implementations of heavy-quark transport and hadronization in the QGP. While most models can describe the data reasonably, some show deviations. 


\section{Summary}
\label{sum}
In these proceedings, the first measurements of D-meson and \Lc-baryon \vtwo performed with ALICE using the 2023 \PbPb data collected in Run 3 were reported. 
In combination with future measurements of \RAA in Run 3, the comparison of D-meson and \Lc-baryon \vtwo will constrain the description of the hadronization process and its impact on \vtwo, allowing for a more precise estimate of the heavy-quark diffusion coefficient. 
Overall, with the inclusion of 2024 and 2025 data, we expect to analyze a data sample with a factor of 3 larger than the current one.
\section*{Acknowledgments}
This work is supported by the \textit{national key research, development program of China} (No. 2024YFA1610800, 2022YFE0116900, and 2018YFE0104700), \textit{National Natural Science Foundation of China} (Grant No. 12175085), and \textit{China Scholarship Council} (Grant No. 202406410012),

\bibliography{template.bib} 
%
%

\end{document}